\documentclass[reprint,prmaterials,notitlepage]{revtex4-2}
\usepackage {graphicx}
\usepackage{multirow}
\usepackage{amsmath}
\begin{document}

\raggedbottom

\title {\bf Nucleation of twinning dislocation loops in fcc metals}

\author {Sweta Kumari and Amlan Dutta }

\affiliation {Department of Metallurgical and Materials Engineering, Indian Institute of Technology Kharagpur,
West Bengal 721302, India}

\date{\today}

\begin{abstract}
Deformation twinning, which occurs in fcc metals only under particular conditions of intrinsic material properties, microstructure, and loading conditions, occupies an indispensable place in their deformation mechanism maps. Nonetheless, dedicated studies have seldom been carried out to explore the fundamental properties of twinning dislocations mediating this process. Here we employ a combination of atomistic computations and continuum modeling to investigate the nucleation of twinning dislocation loops in metals with fcc crystal structure. Besides the conventional layer-by-layer model of twin nucleation, the newly proposed alternate-shear mechanism has been investigated, and the energy barrier and shear stress for twinning loop nucleation have been determined. We find the nucleation stress in the latter mechanism to be smaller than that in the former. The study also highlights the non-uniform variations of the critical loop size and fractional Burgers vector with the applied shear load.
\end{abstract} 

\maketitle

Except for low stacking fault energy (SFE) systems, deformation twinning is typically absent in pure metals and alloys with fcc crystal structure. However, under the non-trivial conditions of loading, the metals with moderate and high SFEs can also exhibit this mechanism. Conditions favorable to deformation twinning include high strain rates and stresses, along with the reduced temperature~\cite{Dynamic_deformation,Macrodeformation,Cu-twin}. In equal channel angular pressing of copper, twinning has been observed in the shear bands representing the regions of severe plastic deformation~\cite{Cu-polycrystal}. In addition to the coarse-grained polycrystalline samples under special loading conditions, nanocrystalline fcc metals like aluminum and copper exhibit deformation twinning even at room temperature and low strain rates~\cite{Al-nanocrystal,Cu-nanocrystal,nanotwinned}. Similarly, nanosized fcc systems like nanoparticles~\cite{nanoparticle}, nanowires~\cite{Al_nanowire}, and nanopillars~\cite{Pt_nanocrystal} also exhibit load-induced twinning.\

Given the elusive nature of deformation twinning, multiple mechanisms have been proposed to explain the formation and growth of twins in fcc metals. While the twin formation in bulk systems is often described in terms of the pole-mechanism, fault-pair model, or layer-by-layer shear model, the shear model is the most feasible mechanism in their nanosized counterparts on account of the confined space available~\cite{FCC_BCC}. Furthermore, Wang\textit{ et al.}~\cite{Wang2017} have recently demonstrated a variation of the conventional layer-by-layer mechanism, which can induce deformation twinning in a very high SFE metal like platinum. Regardless of the exact mechanism in action, the formation of a twin embryo involves plastic slips on three successive \{111\} slip planes, and its growth occurs through subsequent slip over the existing twin boundary. Slip of atomic planes, both during the embryonic and growth stages of the twin-fault, is mediated by the twinning partial dislocations. It underlines the significance of twinning dislocations in the deformation mechanism maps of the fcc metals~\cite{FCC_metals}. A key contribution of the famous Orowan's relationship is establishing a trade-off between the density and velocity of line defects for a given strain rate. Consequently, a comprehensive understanding and quantification of the deformation process remain incomplete unless the mechanism of dislocation nucleation has been explored in detail. In the case of fcc metals, atomistic and line dynamics simulations involving saddle point search have shown remarkable success in computing the activation energies for the nucleation of dislocation loops~\cite{Aubry2011,Jennings2013,Geslin2017}. These computational strategies have been duly complemented with the development of analytical nucleation models. In this direction, the first breakthrough is probably attributable to Gutkin and Ovid’ko~\cite{Gutkin}, who put forward the idea of treating the Burgers vector as a variable non-crystallographic parameter. A significant refinement of this approach was introduced by Aubry\textit{ et al.}~\cite{Aubry2011}, who modified the energetics equation of homogeneous loop nucleation by incorporating the generalized stacking fault energy (GSFE) as an atomistically extracted parameter. Jenning’s\textit{ et al.}~\cite{Jennings2013} further extended this scheme to study the surface nucleation of dislocations in the nanowires of several fcc metals. More recently, Faisal and Weinberger~\cite{power_law} have employed these nucleation models to analyze the power-law exponent that appears in a phenomenological expression of the activation free-energy.\

Despite the improvements in simulation and analytical strategies, the studies on fcc metals have primarily confined to the nucleation of full or Shockley partial dislocation loops~\cite{Aubry2011,Jennings2013,Geslin2017,power_law,Multi-scale}. In contrast, very little information is available regarding the nucleation of twinning dislocations in these systems. It entails a glaring gap in comprehending the deformation mechanism map, which is expected to incorporate the deformation twinning mechanism depending upon the intrinsic characteristics of a material and external loading conditions. In this context, we must note that the slip of atomic planes leading to the formation of a stacking fault is crystallographically distinct from a slip leading to the formation or migration of a twin boundary. It results in a qualitative difference between the GSFE and the generalized planar fault energy (GPFE) of the \{111\} slip plane. In the present study, we explore the homogeneous nucleation of twinning dislocation loops in fcc metals by atomistic simulations. In particular, the nudged elastic band (NEB) method has been employed to measure the activation barrier against the nucleation of twinning loops for two different slip configurations in Ag, Au, and Cu. The choice of these three metals covers a fairly broad range of the intrinsic stacking fault energy, while the two types of slips correspond to the conventional layer-by-layer shear and the newly discovered alternate-shear mechanism. Following the estimation of nucleation paths at different stresses, we compute the generalized twin fault energies for the two configurations considered here. The GPFE acts as a critical material parameter fed to the continuum nucleation model, whose predictive capability has been assessed \textit{vis-\`{a}-vis} the outcome of the NEB calculations.\

\begin{figure}[t]
	\centerline{\includegraphics*[width=8cm, angle=0]{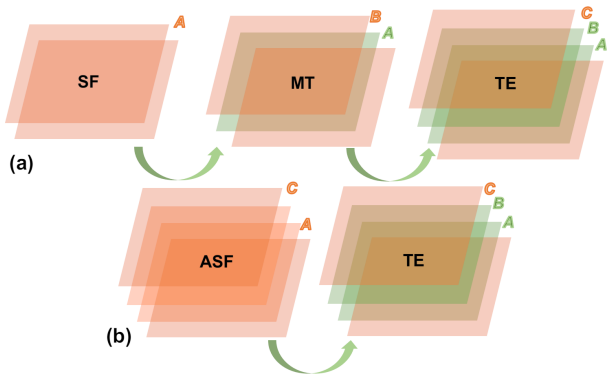}}
	\caption{(color online). Schematic illustration of the (a) layer-by-layer (\textit{ABC}), and (c) alternate-layer (\textit{ACB}) slip configurations. The green layers indicate atomic planes with local fcc packing, whereas the orange ones have local hcp structure. Two adjacent hcp layers constitute an intrinsic stacking fault, while a single layer represents a twin boundary. Subsequent slips over the twin boundaries of the embryo would cause it to grow.}
\end{figure}

Figure 1 schematically demonstrates the two varieties of slip considered in the present work. In the first (Fig. 1(a)), the formation of the twin embryo commences with the formation of an intrinsic stacking fault (SF) when one part of the crystal slips over a \{111\} plane by a relative displacement of $a\langle 11\bar{2} \rangle/6$. A subsequent slip over this stacking fault creates a meta-twin (MT) with two boundaries, and a third subsequent slip over the meta-twin boundary creates the twin embryo (TE). Further slips on the consecutive slip planes would cause the twin boundary to migrate, thereby widening the twin slab. This is the conventional layer-by-layer scheme of deformation twinning, which begins with the \textit{ABC} kind of embryo formation, as depicted in the figure. This scheme has been referred to as Path-A by Wang \textit{et al.} (\textit{c.f.} Fig. 3(a) in Ref.~\cite{Wang2017}). Another possible route is displayed in Fig. 1(b). Here the slips occur on two alternate \{111\} planes, separated by a single plane in between, thereby creating two adjacent intrinsic stacking faults. A third slip on the intermediary plane creates the twin embryo. We can refer to this scheme as the \textit{ACB} shear, which has been designated as Path-B by Wang \textit{et al.} (\textit{c.f.} Fig. 3(b) in Ref.~\cite{Wang2017}) and is very similar to the mechanism reported by Wang and Huang~\cite{Shockley}. On the one hand, the formation of intrinsic stacking faults in both the mechanisms is associated with the glide of Shockley partials with $a\langle 11\bar{2} \rangle/6$ as their Burgers vector. On the other hand, slips leading to the formation and growth of the meta-twin or twin embryo are mediated by the glide of twinning dislocations with the same Burgers vector as that of a Shockley partial.\

The first set of atomistic computations estimates the minimum-energy paths of the nucleation of twinning dislocation loops by employing the climbing-replica version~\cite{NEB} of the NEB method. Here the first configuration involves the loop nucleation over an existing twin boundary and corresponds to the formation and growth of a meta-twin or twin embryo associated with the conventional \textit{ABC} shear mechanism (Fig. 1(a)). The initial structure used in the NEB calculations consists of a crystal of dimensions $16a\langle 11\bar{2} \rangle$ $\times $ $24a\langle\bar{1}10\rangle$ $\times$ $20a\langle 111 \rangle$, with the $x$ and $z$ directions along the Burgers vector and the normal to the slip plane, respectively. The crystal also contains an embryonic twin slab with two twin boundaries. This structure is relaxed to the desired shear stress ($\tau_{xz}$) so that a shear strain is developed. The other end of the replica chain is constructed by creating a twinning shear loop of radius $\sim$10 {\AA} and Burgers vector $a\langle 11\bar{2} \rangle/6$ on the back of the upper twin boundary. Figure 2(a) shows the two ends of the chain of states. The second configuration deals with the nucleation between two adjacent intrinsic stacking faults and corresponds to the twin-forming slip of the \textit{ACB} mechanism shown in Fig. 1(b). Here the initial state of the NEB is similar to that in the first configuration, with the exception of having two adjacent intrinsic stacking faults instead of the twin embryo. The end state consists of a small dislocation loop placed on the \{111\} slip plane between these two stacking faults, as exhibited in Fig. 2(b). Periodic boundary conditions have been imposed in all the three directions.\

\begin{figure}[t]
	\centerline{\includegraphics*[width=8cm, angle=0]{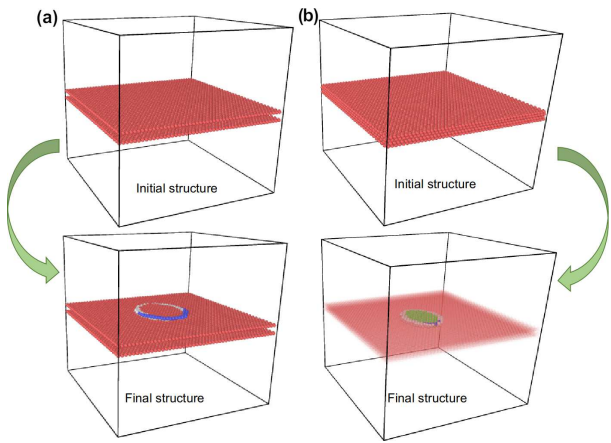}}
	\caption{(color online). Representative structures of the initial and final states for NEB calculations in the (a) layer-by-layer (\textit{ABC}), and (b) alternate-layer (\textit{ACB}) twinning routes. Only the atoms in the twin fault are shown, where the green particles indicate atoms with fcc structure, while the red ones have local hcp environments. The slab of hcp atoms in the ACB configuration is shown as translucent so that the twinning loop at the center of the slab becomes visible. }
\end{figure}

The replica-chain is initialized by creating fourteen intermediate images between the end states through linear interpolation. In order to prevent the loop from completely expanding and sweeping through the slip plane, feedback forces are applied on the end state, which confine it to an isopotential surface. In the NEB computation, the chain of replicas is made to evolve through the recently improved FIRE algorithm~\cite{FIRE_algorithm}, which offers a significant enhancement in performance over its older implementation. To extract the interatomic forces, we select the embedded-atom-model potentials for Au~\cite{Au_potential}, Ag~\cite{Ag_potential}, and Cu~\cite{Cu_potential}, which produce the cohesive energies, SFEs, and stiffness constants in reasonable agreement with those obtained from the DFT calculations.\

The second set of atomistic computations involves quantifying the generalized planar fault energy, which is an input parameter of the nucleation model described further below. Here the set-up is similar to the one created for the initial structure of the NEB calculations, but with free surfaces normal to the $\langle 111 \rangle$ direction. In the first configuration with an embryonic twin, the upper twin boundary is the slip plane, whereas the plane between the two stacking faults is chosen as the slip plane in the second configuration. The part of the crystal above the selected slip plane is displaced laterally relative to the lower part, and the energy of the system is recorded as a function of disregistry. At each step of the lateral displacement, the system is structurally relaxed, such that the atoms are allowed to move only normal to the slip plane. All the atomistic computations reported here have been performed using the LAMMPS molecular dynamics code~\cite{lammps}, whereas the Atomsk~\cite{Atomsk} and OVITO~\cite{Stukowski} codes have been employed for various pre-processing and post-processing tasks, respectively.\

The atomistically informed nucleation model employed in this study is based upon the approach demonstrated by Aubry \textit{et al.}~\cite{Aubry2011}. It expresses the enthalpy of formation of a dislocation loop of radius $r$ at a given stress $\tau$ as,

\begin{equation}
\begin{split}
 E \left( b_{f}, r; \tau \right)=2 \pi r\frac{Gb_{f}^{2}}{8 \pi }\left[\frac{2- \nu }{1- \nu } \left( \ln \frac{8r}{r_{0}}-2 \right) +\frac{1}{2} \right] \\
 + \pi  \left[\gamma _{gpf} \left( b_{f}+u_{0}; \tau \right) - \gamma _{gpf} \left( u_{0}; \tau \right) \right] r^{2}- \pi b_{f} \tau r^{2},
 \end{split}
\end{equation}

\noindent
where $b_{f}$ is a variable fractional Burgers vector and $\gamma(\delta)$ is the GPFE obtained as a function of the disregistry, $\delta$. In the above equation, $u_{0}$ is the offset-displacement calculated by equilibrating the applied stress against the gradient of the GPFE. Apart from the GPFE, the other intrinsic material parameters are the effective shear modulus ($G$) and Poisson's ratio ($\nu$), which in general, depend upon the relevant crystallographic orientation~\cite{Fitzgerald_2010}. $r_{0}$ is the effective cut-off radius, which is the only fitting parameter in this model. Although a somewhat simpler expression for the loop's enthalpy has been employed by Jenning's \textit{et al.}~\cite{Jennings2013}, Eq. (1) is found to yield slightly better results for the systems studied here.\ 

A noteworthy feature of this nucleation model is recognition of the fact that at large enough stress, the GPFE itself becomes sensitive to the applied stress, thereby justifying the incorporation of the stress, $\tau$, as an extrinsic parameter in Eq. (1). Once the stres-dependent GPFE is obtained, the loop-energy is computed over a two-dimensional grid consisting of $b_{f}$ and $r$, and the activation barrier against nucleation of the twinning loop is computed at the saddle point of this energy surface.\

\begin{figure}[b]
	\centerline{\includegraphics*[width=8cm, angle=0]{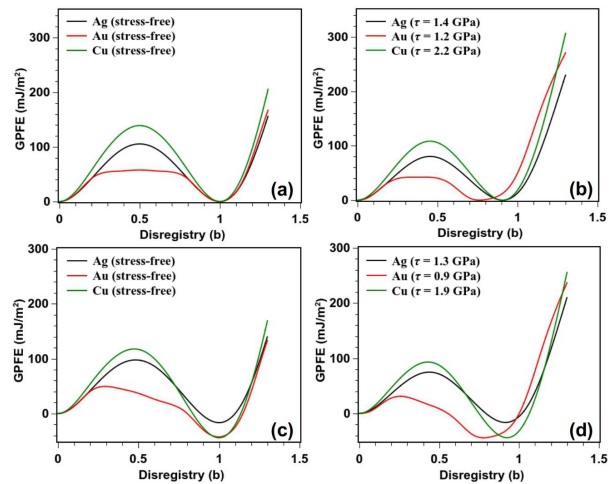}}
	\caption{(color online). Generalized planar fault energies for the (a) stress-free and (b) stressed crystals in the \textit{ABC} shear mechanism. GPFE profiles for the (c) stress-free and (d) stressed crystals in the \textit{ACB} shear model. The indicated shear stress values are the maximum stresses at which the nucleation barriers have been estimated for each of the metals. }
\end{figure}

Figure 3(a) displays the generalized planar fault energies corresponding to the \textit{ABC} shear model for a stress-free crystal, where the slip occurs over a pre-existing twin fault. As recently shown by Pulagam and Dutta~\cite{pulagam2020corestructure}, both the slips illustrated in Fig. 1(a) for the layer-by-layer model and the subsequent slips entailing further thickening of the twin slab, exhibit identical GPFE profiles. This observation is attributed to the equivalent stacking sequence of atomic planes around the fault boundary in each of these slip configurations. Therefore, we present only a single GPFE profile for each metal in Fig. 3(a), which alone represents the planar fault energies for all the slips in the layer-by-layer model.  In a perfect fcc crystal, the slip of $\textbf{b} = a\langle 11\bar{2} \rangle/6$ on the \{111\} plane would produce a stable stacking fault energy, as seen in a typical generalized stacking fault energy profile. In contrast, the GPFE plots in Fig. 3(a) show translational invariance of structural energy at the same disregistry. This is because unlike a perfect fcc crystal, where a slip of $a\langle 11\bar{2} \rangle/6$ creates an intrinsic stacking fault, the slips in the \textit{ABC} shear model merely cause the pre-existing twin boundary to shift by one atomic plane, while its structure and energy remain unaltered. In the presence of an applied shear load, the relative displacements corresponding to the translational invariance in the GPFE profiles get reduced below the original value of $a\langle 11\bar{2} \rangle/6$, as evident from comparing Fig. 3(b) to Fig. 3(a). Moreover, the applied shear load also reduces the unstable twin fault energies, given by the intermediate energy maxima seen in the figures. Both of these stress-effects are attributable to the alteration in atomistic configuration across the slip-plane due to the elastic shear strain.\

As compared to the \textit{ABC} shear, a contrasting feature is visible in the GPFE plots corresponding to slip in the alternate-layer (\textit{ACB}) mechanism. Here we find that the disregistry of $a\langle 11\bar{2} \rangle/6$ leads to a structure with energy, which is lower than the initial energy, and effectively represents a negative stacking fault energy (Fig. 3(c)). This peculiar feature can be understood by realizing that the initial structure with two stacking faults on alternate planes has substantially elevated structural energy, for it is effectively four hcp layers sandwiched between fcc phases. A relative slip of $a\langle 11\bar{2} \rangle/6$ between the two adjacent stacking faults, as illustrated in the third step of Fig. 1(b), causes the middle two layers to return back to their fcc configurations. It explains the drop in the energy of the system with respect to the initial structure. In the thermodynamic viewpoint perceiving an intrinsic stacking fault as a thin hcp phase embedded within the fcc system~\cite{particle_swarm_optimization}, the slip studied in Fig. 3(c) can be perceived as a process that is opposite to that of the formation of a stacking fault in fcc. Similar to what is observed in Fig. 3(b), an applied shear stress affects the GPFE profile in the \textit{ACB} shear model as well (Fig. 3(d)).\

\begin{figure}[t]
	\centerline{\includegraphics*[width=6cm, angle=0]{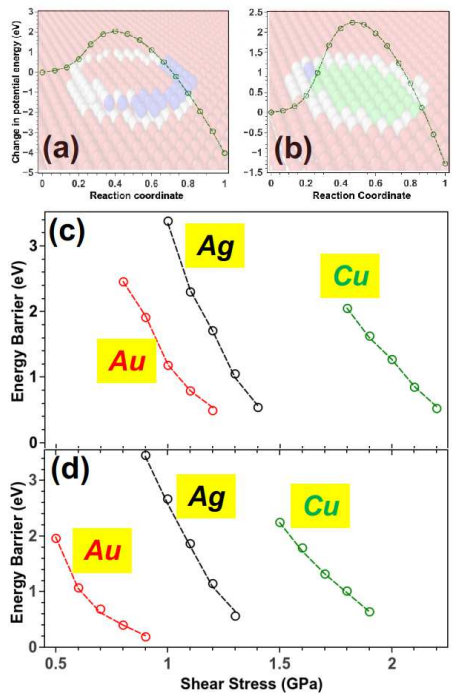}}
	\caption{(color online). Structural energies plotted against the reaction coordinates for twinning loop nucleation in Cu in the (a) layer-by-layer and (b) alternate-shear mechanisms. Energy barriers displayed as the functions of applied shear stress in (c) \textit{ABC} and (d) \textit{ACB} shear models. Circular symbols present the results of NEB calculations, whereas the dashed lines indicate the continuum model's fitted outcomes.}
\end{figure}

Having explored the generalized planar fault energies corresponding to the two shear models, we now examine the transition pathways revealed by the NEB computations. Figures 4(a) and (b) display the representative plots showing the variations in structural energies along the minimum energy paths of twinning loop nucleation in the \textit{ABC} and \textit{ACB} shear models, respectively. Here the critical loop nucleus is easily identified as the replica with the largest potential energy, and the corresponding snapshots of the critical twinning loops are also displayed as the translucent backgrounds. The energy barriers are computed at different stresses for the layer-by-layer shear mechanism and are exhibited in Fig. 4(c). We find that the stress required for a given energy barrier is largest for Cu and smallest for Au. This trend can be intuitively understood by realizing that the unstable twin fault energy dominates the mechanism of loop nucleation modeled in Eq. (1), which follows the order Cu $>$ Ag $>$ Au (Figs. 3(a) and (b)).\

In the case of a shear loop of Shockley partial, a positive stacking fault energy tends to oppose its expansion. Therefore, it is interesting to note that the negative fault energy observed in Figs. 3(c) and (d) is expected to support the expansion of the loop in the \textit{ACB} nucleation model. Further, the unstable fault energies are smaller than their corresponding counterparts in the layer-by-layer model. This observation leads us to predict reduced stresses for loop nucleation. The critical loop nucleus in the alternate-layer mechanism, as seen in Fig. 4(b), reveals that unlike the loop in the \textit{ABC} model, where the atoms inside the loop have the same twin-boundary structure (Fig. 4(a)), the atoms inside the loop in the \textit{ACB} model have local fcc structure in contrast to the hcp atoms outside it. Figure 4(d) displays a reduction in the nucleation barrier with an increase in the applied shear stress. A direct comparison with Fig. 4(b) shows that the nucleation stresses of twinning dislocation loops in all the three metals are indeed smaller than those in the \textit{ABC} model, as expected from the GPFE profiles. \

\begin{figure}[t]
	\centerline{\includegraphics*[width=6cm, angle=0]{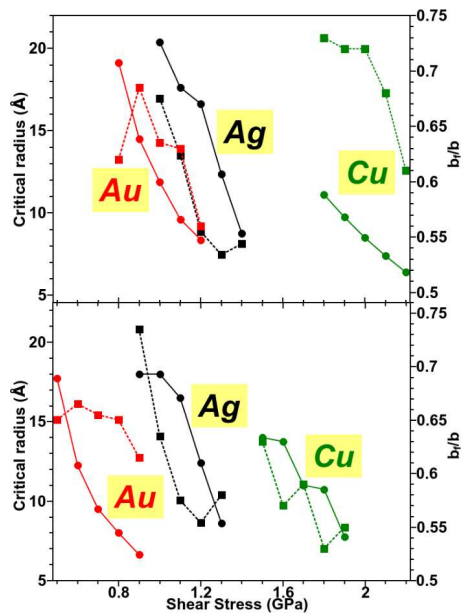}}
	\caption{(color online). Radii (circular symbols) of the critical twinning loops and associated fractional Burgers vectors (square symbols) plotted against the applied shear loads in the (a) layer-by-layer and (b) alternate-shear mechanisms. To facilitate comparison among the three metals, the fractional Burgers vector ($b_f$) is given as fraction of the corresponding crystallographic Burgers vector, $b = a/\sqrt{6}$.}
\end{figure}

Equation (1) indicates that besides yielding the energy barrier of twinning loop nucleation, the saddle point along the transition pathway also reveals the fractional Burgers vector, $b_f$, and the critical loop size, $r$. Figure 5(a) displays these critical parameters of the nucleated twinning dislocation loops as calculated for the \textit{ABC} shear model, whereas Fig. 5(b) presents the same for the \textit{ACB} mechanism. It is pertinent to recall that in the conventional model of loop nucleation, the Burgers vector is assumed to be fixed and the same as the crystallographic Burgers vector, and predicts a uniform reduction in the critical radius with an increase in applied shear stress. However, the results in Fig. 5 underline the implications of replacing the crystallographic Burgers vector ($b$) by its fractional counterpart ($ b_f $). We note that in both the shear models, the variation of loop radius with applied load is not essentially uniform and can even be non-monotonic. For instance, in the \textit{ACB} shear, the critical loop radius in Ag at 1 GPa is slightly larger than that at 0.9 GPa (Fig. 5(b)). Such a trend can be understood by observing that the fractional Burgers vector's variation with increasing applied stress is highly non-uniform and unpredictable. In the example mentioned above, we find that a substantial drop in $b_{f}$ from the stress of 1 GPa to 0.9 GPa causes the loop radius to increase instead of decrease, as opposed to our intuitive expectation. Such results firmly imply that the conventional, text-book model of loop nucleation is not only quantitatively inaccurate but can be inadequate in predicting the qualitative trend as well.\

By way of summary, this study examines the nucleation of twinning dislocation loops in fcc metals through the technique of transition-pathway-search and atomistically informed continuum model. In particular, we consider the newly proposed alternate-shear model of twin formation, in addition to the conventional layer-by-layer model. A particularly interesting feature of the twinning dislocations is that while their Burgers vectors resemble those of Shockley partials, the generalized fault energy in the layer-by-layer nucleation does not exhibit the stable stacking fault energy, which is otherwise a quintessential characteristic of the GSFE. Besides, the alternate-shear model reveals a negative stacking fault energy, which motivates a comparison between the twinning loop nucleation in these two mechanisms. Our computations indicate that the nucleation stress of twinning loops is larger for metal with higher unstable fault energy. Furthermore, the nucleation stress for a given metal is smaller in the alternate-shear mechanism, as compared to its layer-by-layer counterpart. When tuned with a single free parameter, the continuum model produces excellent fits to the NEB results. A closer look at the critical loop reveals that its fractional Burgers vector plays a vital role in dictating the critical loop size and stress of nucleation.\

The computational results highlight the significance of perceiving the twinning deformation in terms of nucleation and growth of twinning partial loops in the fcc materials. An obvious extension of this work is to study the process of heterogeneous nucleation of twinning loops. Although the surface nucleation would be comparatively easier to model analytically due to the simpler treatment of image stresses, it would be more challenging to model grain boundary nucleation. Moreover, as the computations provide us with the energy barrier to loop nucleation, it can be subsequently utilized to estimate the kinetic parameters associated with the process at different strain rates and temperatures. The results and insights rendered here can motivate further investigations into the fundamental computational modeling of twinning dislocations.\\

We acknowledge the Department of Science and Technology, Govt. of India, for partially funding the present work under the ECRA scheme. We also thank the Indian Institute of Technology Kharagpur, for financial support through the ISIRD grant.

%

\end{document}